\documentclass[runningheads]{llncs}
\usepackage[T1]{fontenc}
\usepackage{multirow}
\usepackage{graphicx}
\usepackage[pagebackref=true,breaklinks=true,colorlinks,bookmarks=false]{hyperref}

\usepackage{breakurl}
\usepackage{url}

\usepackage{cite}
\usepackage{amsmath,amssymb,amsfonts}
\usepackage{graphicx, color}
\usepackage{subcaption}
\usepackage[export]{adjustbox}
\usepackage{textcomp}
\usepackage{url} 
\usepackage{makecell}
\usepackage[table]{xcolor} 
\usepackage{paralist}
\usepackage{enumitem}
\usepackage{multirow}
\usepackage{hyperref}
\usepackage{algorithmic}
\usepackage{xcolor}

\usepackage{graphicx}

\usepackage{caption}
\usepackage{subcaption}
\usepackage{wrapfig}

\usepackage{orcidlink}

\begin{document}
\title{Challenge Summary: U-MedSAM Uncertainty-aware MedSAM for Medical Image Segmentation}
\titlerunning{Abbreviated paper title}
%


\author{
Xin Wang\inst{1}  \orcidlink{0000-0002-7528-2407} \and
Xiaoyu Liu\inst{2} \and
Peng Huang\inst{3} \and
Pu Huang\inst{2} \and \\
Shu Hu\inst{4} \and 
Hongtu Zhu\inst{5}
} 
%
\authorrunning{Xin Wang et al.}
%
\institute{
College of Integrated Health Sciences and the AI Plus Institute, University at Albany, State University of New York (SUNY) \and
School of Physics and Electronics, Shandong Normal University \and
School of Computing and Artificial Intelligence, Southwest Jiaotong University \and
Department of Computer and Information Technology, Purdue University \and
University of North Carolina at Chapel Hill
\\
\email{
\{xwang56\}@albany.edu,
\{liuxy2823\}@163.com,  
\{huangpeng\}@my.swjtu.edu.cn,
\{huangpu\}@sdnu.edu.cn, 
\{hu968\}@purdue.edu,
\{htzhu\}@email.unc.edu
}
}
\maketitle              
\begin{abstract}
Medical Image Foundation Models have proven to be powerful tools for mask prediction across various datasets. 
However, accurately assessing the uncertainty of their predictions remains a significant challenge. 
To address this, we propose a new model, U-MedSAM, which integrates the MedSAM model with an uncertainty-aware loss function and the Sharpness-Aware Minimization (SharpMin) optimizer.
The uncertainty-aware loss function automatically combines region-based, distribution-based, and pixel-based loss designs to enhance segmentation accuracy and robustness. 
SharpMin improves generalization by finding flat minima in the loss landscape, thereby reducing overfitting. 
Our method was evaluated in the CVPR24 MedSAM on Laptop challenge, where U-MedSAM demonstrated promising performance.

\end{abstract}

\section{Introduction}


\begin{wrapfigure}{R}{0.5\textwidth}
\vspace{-2mm}
\centering
    \includegraphics[trim=1 1 1 1, clip,keepaspectratio, width=0.5\textwidth]{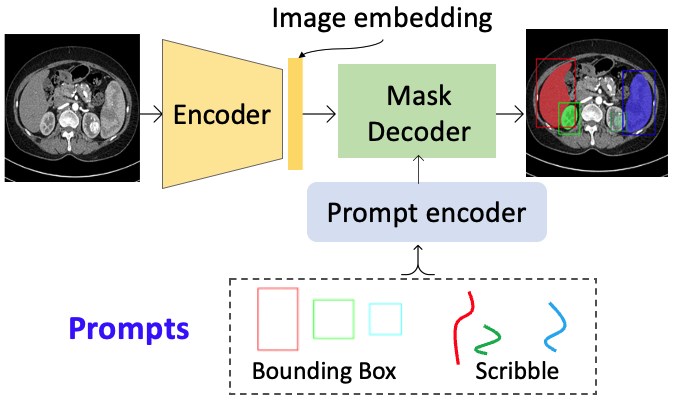}
  \caption{\small \it Overview of the MedSAM \cite{MedSAM}.}
\vspace{-3mm}
\label{fig_th1}
\end{wrapfigure}

Medical image segmentation is vital in clinical practice, enabling precise medicine \cite{wang2024artificial,wang2023deep}, assessing therapeutic outcomes, and disease diagnosis by delineating organ boundaries and pathological regions, enhancing anatomical understanding, and abnormality detection \cite{huang2024robustly}.
Early segmentation models for medical images are typically designed for specific tasks with limited data. 
They might not handle the complex patterns and minute variations in medical images, which are often critical for clinical diagnosis and health science study \cite{lin2024robust}.

Recently, foundation models, such as MedSAM~\cite{MedSAM} have been developed for medical image segmentation \cite{SAM-ICCV23} \cite{MobileSAM} \cite{EfficientViT-SAM} (See Fig. \ref{fig_th1}), which are more strong, efficient, and applicable to different data modalities and situations. 
However,  the uncertainty of the learning process associated with MedSAM on the diverse dataset has not been thoroughly investigated. 
By understanding this uncertainty, we can improve the robustness of MedSAM, making it more trustworthy for practical applications \cite{tsai2024uu}.



%
%

In this paper, we address the above challenge by proposing the U-MedSAM model, an extension of the MedSAM model~\cite{MedSAM}, and improve its training by incorporating a {\bf novel uncertainty-aware hidden task learning}. 
The uncertainty-aware framework enables the model to balance different aspects of the segmentation sub-tasks (e.g., boundary, pixel, region) adaptively, improving overall performance and robustness, and addressing issues like class imbalance\cite{cho2022}. 
Specifically, our newly introduced uncertainty-aware loss function integrates three components \cite{kendall2018multi}: 
(1) {\em Pixel-based loss}: Measures differences at the pixel level ({\em e.g.}, Mean Squared Error loss) \cite{LossOdyssey}.
(2) {\em Region-based loss}: Used for regions segmentation ({\em e.g.}, Dice coefficient loss). 
(3) {\em Distribution-based loss}: Compares predicted and ground truth distributions ({\em e.g.}, Cross-entropy loss and Shape-Distance loss). 
Finally, we utilize the {\bf Sharpness-Aware Minimization (SharpMin) optimizer} \cite{harper2022,lin2024robust,lin2024robust2,lin2024robust3,lin2024preserving} to improve the generalization of our U-MedSAM.
We evaluate and compare our method against state-of-the-art MedSAM models, demonstrating superior performance. 
Our approach achieves optimal results by promoting sharper and more precise segmentation boundaries, thereby enhancing the accuracy and robustness.

In summary, we make the following key contributions:

\begin{itemize}[leftmargin=10pt] \itemsep -.1em
    \item We introduce an uncertainty-aware learning loss combining pixel-based loss ({\em e.g.}, Mean Squared Error loss), region-based loss ({\em e.g.}, Dice coefficient loss), and distribution-based loss ({\em e.g.}, Cross-entropy loss and Shape-Distance loss). This design enables the model to conveniently incorporate any loss function for training the model jointly.
    
    \item By employing {\bf auto-learnable weights}~\cite{kim2023} rather than fixed weights, our model can dynamically adjust the contribution of each loss component based on the uncertainty associated with each prediction. 
    This approach allows the model to prioritize confident predictions and reduce the impact of ambiguous or noisy data. 
    
    \item By using the Sharpness-Aware Minimization (ShapMin) optimizer, our U-MedSAM finds parameter values that result in flat minima in the loss landscape. 
    This technique improves the model's ability to generalize and mitigates overfitting, a common issue in deep learning applications for medical imaging~\cite{murray2022}. 

\end{itemize}

\section{Method}

\begin{figure*}[t]
\centering
\includegraphics[width=0.99\textwidth]{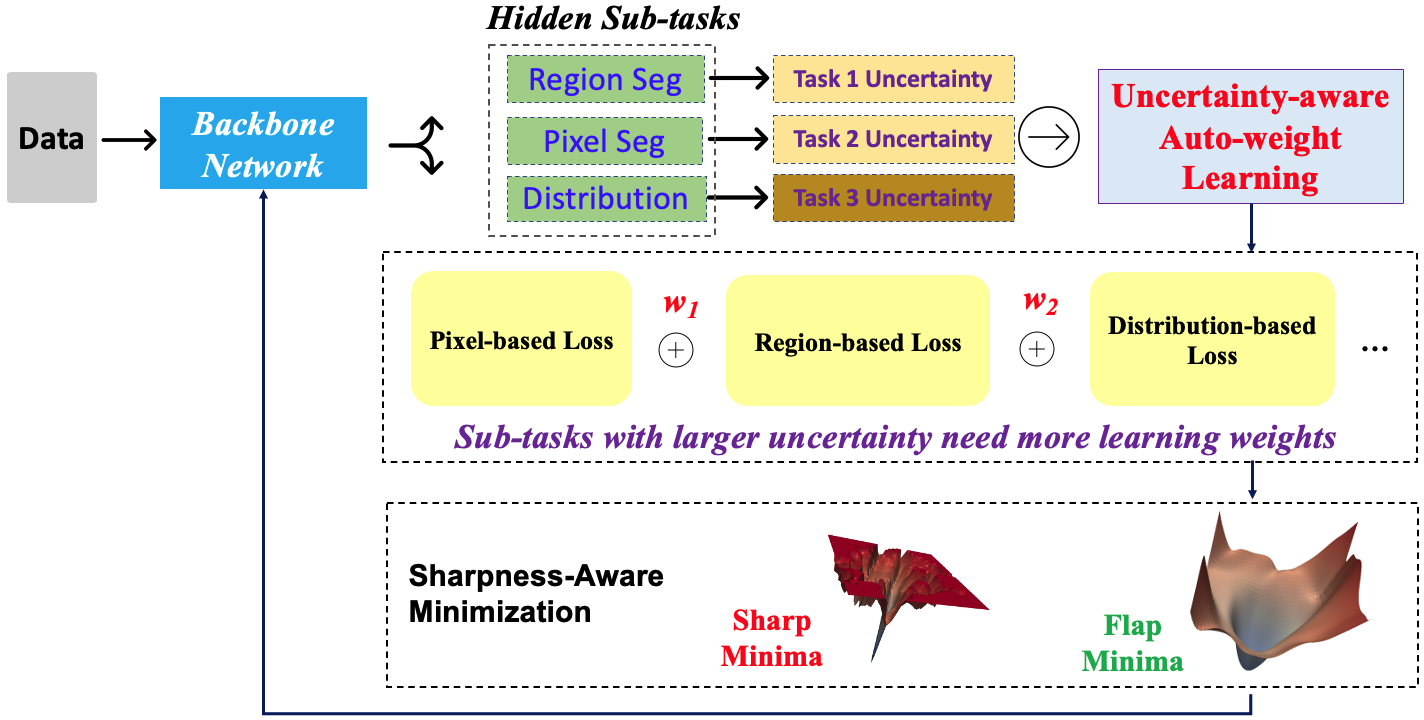}
\caption{Overview of our proposed {\bf U-MedSAM model}: Building on the MedSAM architecture, we encode input images and introduce a novel uncertainty-aware learning method for automatic weight learning across multiple loss functions. 
The Sharpness-Aware Minimization (SharpMin) optimizer is employed, operating within a flattened loss landscape to enhance the generalization.
}
\label{fig:freamwork}
\end{figure*}

This section will describe the entire workflow and implementation details. Fig.~\ref{fig:freamwork} shows an overview of our proposed method. 

\subsection{Preprocessing}



We follow the preprocessing in LiteMedSAM implementation (baeline) \cite{MedSAM}. 



\subsection{Proposed Method}



\subsubsection{Knowledge Distillation}
In the LiteMedSAM model, both the prompt encoder and mask decoder are lightweight models, and the image encoder compresses the model through TinyViT. However, TinyViT uses Softmax-based self-attention mechanism, and the computational complexity is still high. So, in order to reduce the computational complexity of the model, we use EfficientViT to replace TinyViT. EfficientViT reduces the computational complexity from quadratic to linear by lightweight ReLU linear attention. During the initialisation process, we use TinyViT as a teacher model for knowledge distillation, efficiently migrated to EfficientViT. Prompt encoder and mask decoder are already very lightweight models and no other changes were made. The knowledge distillation process was supervised using the L2 loss function.

\subsubsection{Uncertainty-aware Auto-learning.}
\label{sec:uncertainty-aware:loss}

Uncertainty-aware learning enables the model to adapt its learning process based on detected uncertainties, enhancing resilience and precision by focusing on confident predictions and minimizing the influence of ambiguous ones \cite{kendall2018multi}. 
This optimization of the training process across various datasets results in improved overall performance \cite{zhao2020uncertainty,zhao2019uncertainty}.

By incorporating uncertainty into the loss computation, the model can dynamically assign weights to each loss component ~\cite{hu2023rank}. This strategy effectively balances global and local accuracy while mitigating the impact of class imbalance. 
This approach allows the model to prioritize learning from reliable instances and reduce the impact of potentially erroneous or ambiguous data. Consequently, the model becomes more resilient to noisy or ambiguous data, leading to significantly improved segmentation performance~\cite{hu2023umednerf,wang2024neural,peng2024uncertainty}.

The uncertainty-aware loss $\mathcal{L}$ if formulated as following:


\begin{equation}
    \begin{aligned}
        \mathcal{L} = \sum_{m=1}^{M} \left( \frac{1}{2\sigma_m^2} \mathcal{L}_m + \log(1 + \sigma_m^2) \right)
    \end{aligned}
    \label{eq:uncertainty-aware_loss}
\end{equation}

where \( M \) is the number of individual loss components, \( \mathcal{L}_m \) represents each individual loss component (such as DC, CE, MSE, and Shape-Distance loss), and \( \sigma_m \) are learnable noise parameters. 
Larger noise levels indicate that the talk is not properly understood, necessitating greater learning weights.
These learnable parameters are optimized during the training process to minimize the overall loss. The last term can constrain the noise to be increased too much \cite{tsai2024uuj}.  
Specifically, we incorporate the following loss functions, see the survey \cite{jadon2020survey} to find more losses.

\begin{enumerate}[leftmargin=12pt]  \itemsep -.1em

    \item {\bf Mean Squared Error (MSE) loss}: This pixel-based loss ensures accurate classification of individual pixels by calculating the squared difference between each pixel's predicted value and the true label to quantify the model's prediction accuracy~\cite{jadon2020survey}.

    \item {\bf Dice Coefficient (DC) loss}: This region-based metric emphasizes the overlap between predicted and ground truth areas, preserving the accuracy of the shape and boundaries of segmented regions~\cite{jadon2020survey}.

    \item {\bf Cross-Entropy (CE) loss}: This distribution-based loss ensures accurate categorization of individual pixels, improving classification precision~\cite{jadon2020survey}.
        
    \item {\bf Shape-Distance (SD) loss}: This distribution-based loss enhances shape features in image segmentation, strengthens boundary alignment, and ensures that the model pays closer attention to the geometric and structural information of the targets during the training process~\cite{chalcroft2023large}.

\end{enumerate}

\subsubsection{Sharpness-Aware Minimization.}
\label{sec:SAM:opt}
Our approach employs Sharpness aware Minimization (SharpMin) optimization \cite{harper2022,tsai2024uu,lin2024ai,lin2024robust3,lin2024robust,lin2024preserving} to enhance the generalization ability of U-MedSAM. 
By flattening the loss landscape, Sharpness-Aware optimization improves the model's generalization potential. 
Traditional optimization techniques target the lowest points in the loss landscape, which can be steep and result in poor generalization on new data. 
In contrast, Sharpness-Aware Minimization identifies smoother minima, regions in the parameter space where the model's performance remains consistent and less susceptible to disturbances.

\begin{figure*}[h]
\centering
\includegraphics[width=0.8\textwidth]{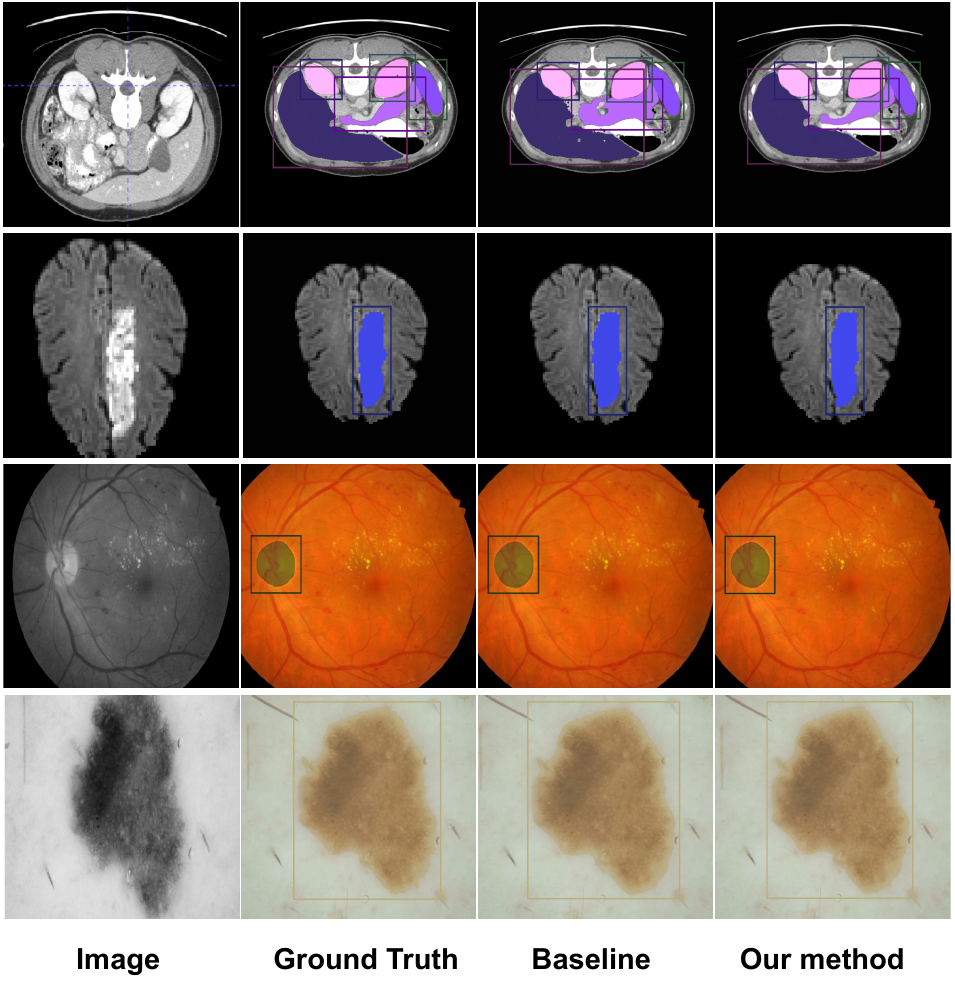}
\vspace{-3mm}
\caption{Qualitative results of well-segmented from various modalities. 
}
\vspace{-3mm}
\label{fig:goodcase}
\end{figure*}

\subsection{Post-processing}


No post-processing techniques were applied, and no strategies were implemented to enhance inference speed.


\section{Experiments}
\subsection{Dataset and evaluation measures}
We only used the challenge dataset for model development.

The evaluation metrics consist of two accuracy measures: the Dice Similarity Coefficient (DSC) and Normalized Surface Dice (NSD), along with an efficiency measure of running time. DSC is used to measure the degree of overlap between the segmentation result and the ground truth. And NSD is a surface-based distance metric for measuring the difference between segmentation boundaries, which takes into account the distance between the boundary of the segmentation result and the label boundary. They are calculated as follows in equations~\ref{eq:DSC} and~\ref{eq:NSD}:
\begin{equation}
\label{eq:DSC}
    \mathrm{DSC}=\frac{2\times|y\cap \hat{y}|}{|y|+|\hat{y}|}
\end{equation}
\begin{equation}
\label{eq:NSD}
    \mathrm{NSD}=\frac{|\{x\in y:d(x,\hat{y})\leq T\}|+|\{x\in \hat{y}:d(x,y)\leq T\}|}{|y|+|\hat{y}|}
\end{equation}
where, $y$ and $\hat{y}$ represent the segmentation result and the ground truth, respectively. $d(x,A)$ denotes the shortest distance from point $x$ to the surface of set $A$, and $T$ is the tolerance distance threshold. Both DSC and NSD are closer to 1, indicating better segmentation performance, which implies a higher degree of overlap and boundary similarity between the predicted segmentation and the ground truth annotation.
Together, these metrics contribute to the overall ranking computation.
\begin{figure*}[t]
\centering
\includegraphics[width=0.8\textwidth]{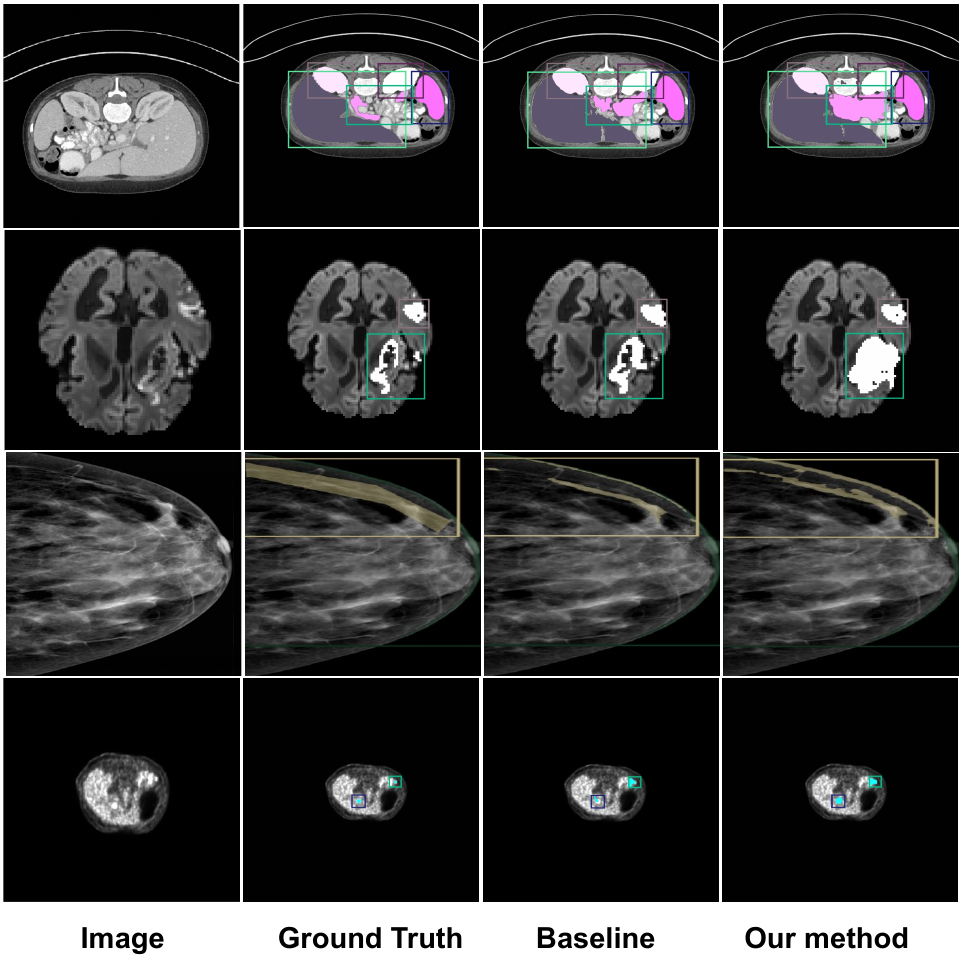}
\vspace{-3mm}
\caption{Qualitative results of challenging examples from various modalities. 
}
\vspace{-3mm}
\label{fig:badcase}
\end{figure*}

\subsection{Implementation details}
\subsubsection{Environment settings}
The development environments and requirements are presented in Table~\ref{table:env}.

Stage 1. Code and models are available at:
\url{https://github.com/liangzw599/Co-developed-by-LiteMedSAM}.

Stage 2 (Post challenge analysis). Code and models are available at: \url{https://github.com/liangzw599/Co-developed-by-LiteMedSAM}.



\begin{table}[t]
\caption{Development environments and requirements. }\label{table:env}
\centering
\begin{tabular}{ll}
\hline
System       &  Ubuntu 20.04.1 LTS\\
\hline
CPU   & Intel(R) Xeon(R) CPU E5-2690 v4 @ 2.60GHz \\
\hline
RAM                         &125GB; 1.03MT$/$s\\
\hline
GPU (number and type)                         & Four NVIDIA Titan Xp\\
\hline
CUDA version                  & 11.4 \\                          \hline
Programming language                 & Python 3.10.14\\ 
\hline
Deep learning framework & torch 2.1.2, torchvision 0.16.2 \\
\hline
\end{tabular}
\end{table}

\begin{table*}[t]
\caption{Training protocols. }
\label{table:training}
\begin{center}
\scalebox{0.91}{
\begin{tabular}{ll} 
\hline
Pre-trained Model         & SAM~\cite{SAM-ICCV23} MedSAM~\cite{MedSAM} \\
\hline
Batch size                    & 2 \\
\hline 
Patch size & 256$\times$256$\times$3  \\ 
\hline
Total epochs & 135 \\
\hline
Optimizer          &   AdamW     \\ \hline
Initial learning rate (lr)  & 0.00005 \\ \hline
Lr decay schedule & ReduceLROnPlateau (mode = min,factor = 0.9,\\  & patience = 5, cooldown = 0) \\  \hline
Training time                                           & 370 hours \\  \hline 
Loss function &  Dice loss, BCE loss, SD loss and IoU loss    \\  \hline 
Number of model parameters    & N/A - not tracked \\ \hline
Number of flops & N/A - not tracked \\ \hline
\end{tabular}
}
\end{center}

\end{table*}

\subsubsection{Training protocols} We follow the training protocols in LiteMedSAM implementation (baseline) \cite{MedSAM}.


\section{Results and discussion}





\begin{figure*}[h]
\centering
\includegraphics[width=0.8\textwidth]{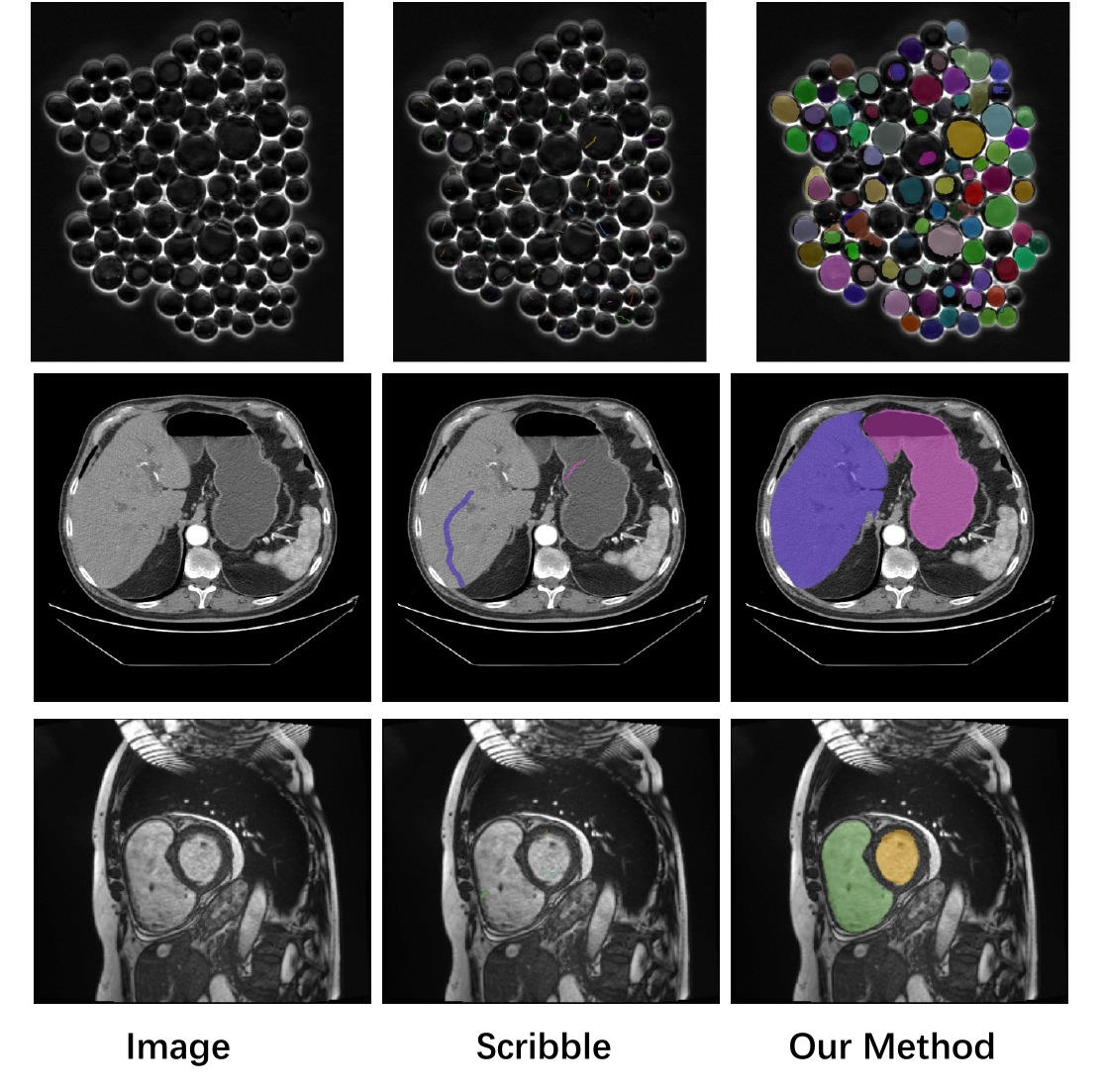}
\vspace{-3mm}
\caption{Scribble-based cases with good results for various modality segmentations. 
}
\vspace{-3mm}
\label{fig:scribble_good_case}
\end{figure*}

\begin{table}[t]
\caption{Quantitative evaluation results. 
}\label{tab:final-results}
\centering
\scalebox{0.9}{
\begin{tabular}{l|cc|cc|cc|cc}
\hline
\multirow{2}{*}{Target} & \multicolumn{2}{c|}{LiteMedSAM} & \multicolumn{2}{c|}{Only SD-loss} & \multicolumn{2}{c|}{No SharpMin} & \multicolumn{2}{c}{U-MedSAM} \\ \cline{2-9} 
                        & DSC(\%)       & NSD(\%)       & DSC(\%)           & NSD(\%)           & DSC(\%)           & NSD (\%)          & DSC(\%)      & NSD (\%)      \\ \hline
CT                      & 89.10         & 91.03         & 87.20             & 89.63             & 89.61                   & 91.66                   & \textbf{89.89}              & \textbf{91.69}               \\
MR                      & \textbf{83.28}         & \textbf{86.10}         & 79.27             & 82.01             & 81.68                   & 83.81                   & 82.79              & 84.61               \\
PET                     & 55.10         & 29.12         & 65.23             & 47.23             & 66.34                   & \textbf{50.17}                   & \textbf{66.63}              & 48.71               \\
US                      & \textbf{94.77}         & \textbf{96.81}         & 83.40             & 88.63             & 84.30                   & 89.27                   & 85.23              & 90.38               \\
X-Ray                   & 75.83         & 80.39         & 78.59             & 84.31             & 83.22                   & 88.94                   & \textbf{85.28}              & \textbf{90.14}               \\
Dermotology             & 92.47         & 93.85         & 93.83             & 95.32             & 93.89                   & 95.34                   & \textbf{94.10}              & \textbf{95.58}               \\
Endoscopy               & 96.04         & 98.11         & 93.60             & 96.24             & 95.65                   & 97.93                   & \textbf{96.24}              & \textbf{98.33}              \\
Fundus                  & 94.81         & 96.41         & 95.45             & 97.05             & 94.87                   & 96.50                   & \textbf{95.66}              & \textbf{97.26}               \\
Microscopy              & 61.63         & 65.38         & 78.20             & 84.59             & \textbf{79.92}                   & \textbf{86.64}                   & 79.05              & 85.33              \\ \hline
Average                 & 82.56         & 81.91         & 83.86             & 85.00             & 85.48                   & 86.64                   & \textbf{86.10}              & \textbf{86.89}               \\ \hline
\end{tabular}
}
\end{table}

\begin{table}[t]
\caption{Quantitative evaluation of segmentation efficiency in terms of running time (s). Note: The inference process cannot use GPU. 
}
\centering
\begin{tabular}{lccccc}
\hline
Case ID &  Size  & Num. Objects    & Baseline & Proposed \\ \hline
3DBox\_CT\_0566    & (287, 512, 512)  & 6 & 376.4      & 286.3     \\
3DBox\_CT\_0888    & (237, 512, 512)  & 6 & 100.5      & 88.7               \\
3DBox\_CT\_0860    & (246, 512, 512)  & 1 &17.7        & 18.5              \\
3DBox\_MR\_0621    & (115, 400, 400)  & 6 &157.1       & 146.7              \\
3DBox\_MR\_0121    &  (64, 290, 320)  & 6 & 99.9       & 73.5             \\
3DBox\_MR\_0179    & (84, 512, 512)  & 1  & 17.1       & 9.8             \\
3DBox\_PET\_0001	& (264, 200, 200)  & 1  & 12.1     & 13.2             \\
2DBox\_US\_0525	& (256, 256, 3)  & 1    &    6.3       & 5.4            \\
2DBox\_X-Ray\_0053 & (320, 640, 3)  & 34    &    7.3   & 6.5            \\
2DBox\_Dermoscopy\_0003    & (3024, 4032, 3)  & 1    &6.5    & 3.2               \\ 
2DBox\_Endoscopy\_0086  & (480, 560, 3)  & 1    &    6.1 & 7.3            \\ 
2DBox\_Fundus\_0003	& (2048, 2048, 3)  & 1    &    6.1  & 7.5          \\
2DBox\_Microscope\_0008	& (1536, 2040, 3)  & 19    &    6.8 & 6.4    \\
2DBox\_Microscope\_0016    & (1920, 2560, 3)  & 241    &   19.1 & 15.9             \\
\hline
\end{tabular}
\label{runtime}
\end{table}


\subsection{Quantitative results on validation set}

Table~\ref{tab:final-results} summarizes these results. The results demonstrate that incorporating uncertainty-aware loss and the SharpMin optimization indeed improves segmentation accuracy.

Training using only the SD loss yields 83.86\% DSC. 
In comparison, training using the uncertainty-aware loss (which incorporates the DC loss, CE loss, and focused loss), enhances DSC to 85.48\%. 
This result shows the capability of the uncertainty-aware loss to increase the robustness and accuracy of models by prioritizing confident predictions and minimizing the impact of uncertain ones. 
It effectively addresses the limitations of using a single loss function by balancing various aspects of the segmentation task.

Incorporating the SharpMin optimization into the uncertainty-aware loss yields 86.10\% DSC, the highest among the examined approaches. SharpMin improves segmentation accuracy by generating clearer and more accurate borders, guiding the optimization process toward flat minima in the loss landscape. This leads to better generalization across diverse data samples. The increase in DSC from 85.48\% to 86.10\% highlights the effectiveness of SharpMin in enhancing segmentation results and more precisely identifying heart components.

\subsection{Qualitative results on validation set}
\begin{figure*}[h]
\centering
\includegraphics[width=0.8\textwidth]{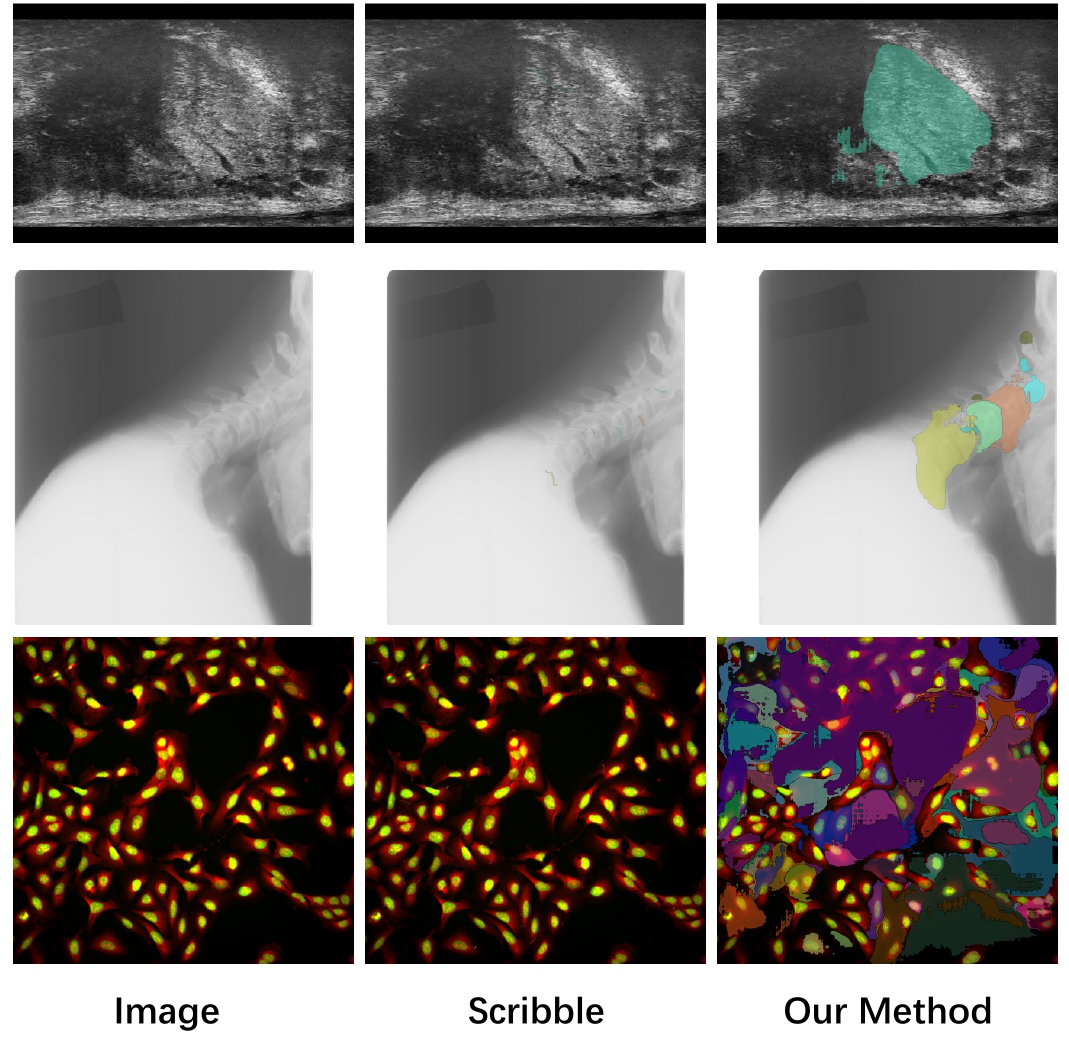}
\vspace{-3mm}
\caption{Scribble-based cases with bad results for various modality segmentations. 
}
\vspace{-3mm}
\label{fig:scribble_bad_case}
\end{figure*}

We show some examples with good segmentation results in Fig. \ref{fig:goodcase} and examples with bad segmentation results in Fig. \ref{fig:badcase}. Fig.~\ref{fig:scribble_good_case} and Fig.~\ref{fig:scribble_bad_case} show the results of good and poor visualisation of scribble-based segmentation, respectively.



\subsection{Segmentation efficiency results on validation set}

As shown in Table \ref{runtime}, we compared the running speed of the proposed method to the LiteMedSAM (baseline) \cite{MedSAM}.

\subsection{Results on final testing set}
This is a placeholder. We will announce the testing results for Post-challenge analysis.

\subsection{Limitation and future work}

Despite achieving promising dice scores, we did not optimize the model's speed, encountering limitations in this area. 

In future work, we will explore various model compression techniques to address these speed constraints. 
Techniques such as quantization and tensor compression will be investigated to enhance the model's efficiency. 
Additionally, we will examine other advanced methods and approaches to further improve the performance and speed of the model.
Our goal is to balance high accuracy with faster processing times, ensuring the model's effectiveness and efficiency for practical applications. By integrating these enhancements, we aim to develop a more robust and versatile model capable of handling real-world scenarios with greater ease and reliability.

\section{Conclusion}

We present U-MedSAM, an innovative model engineered for robust medical image segmentation. 
This model combines the MedSAM architecture with an uncertainty-aware learning framework to dynamically adjust the contribution of multiple loss functions. 
By employing the SharpMin optimizer, U-MedSAM is steered towards flat minima in the loss landscape, enhancing its resilience and improving generalization. 
Comparative analysis with top-performing models highlights U-MedSAM's superior accuracy and robustness in segmentation performance.

\subsubsection{Acknowledgements} We thank all the data owners for making the medical images publicly available and CodaLab~\cite{codabench} for hosting the challenge platform. 

Xin Wang is supported by the University at Albany Start-up Grant.

\subsubsection{Disclosure of Interests} The authors have no competing interests to declare that are relevant to the content of this article. 

%
%
%
\bibliographystyle{splncs04}
\bibliography{ref}

\newpage
\begin{table}[!htbp]
\caption{Checklist Table. Please fill out this checklist table in the answer column.}
\centering
\begin{tabular}{ll}
\hline
Requirements                                                                                                                    & Answer        \\ \hline
A meaningful title                                                                                                              & Yes        \\ \hline
The number of authors ($\leq$6)                                                                                                             & 6        \\ \hline
Author affiliations and ORCID                                                                                           & Yes        \\ \hline
Corresponding author email is presented                                                                                                  & Yes        \\ \hline
Validation scores are presented in the abstract                                                                                 & Yes        \\ \hline
\begin{tabular}[c]{@{}l@{}}Introduction includes at least three parts: \\ background, related work, and motivation\end{tabular} & Yes        \\ \hline
A pipeline/network figure is provided                                                                                           & Figure 2 \\ \hline
Pre-processing                                                                                                                  & Page 3  \\ \hline
Strategies to data augmentation                                                                                             & Page 3   \\ \hline
Strategies to improve model inference                                                                                           & Page 4   \\ \hline
Post-processing                                                                                                                 & Page 4   \\ \hline
Environment setting table is provided                                                                                           & Table 1  \\ \hline
Training protocol table is provided                                                                                             & Table 2  \\ \hline
Ablation study                                                                                                                  & Page 6   \\ \hline
Efficiency evaluation results are provided                                                                                     & Table 4 \\ \hline
Visualized segmentation example is provided                                                                                     & Figure 3,4 \\ \hline
Limitation and future work are presented                                                                                        & Yes        \\ \hline
Reference format is consistent.  & Yes        \\ \hline
Main text >= 8 pages (not include references and appendix)  & Yes        \\ \hline

\end{tabular}
\end{table}

\end{document}